\documentclass[
aps,
prb,
twocolumn,
superscriptaddress,
 amsmath,amssymb,
]{revtex4-2}

\usepackage{graphicx}
\usepackage{dcolumn}
\usepackage{bm}
\usepackage{txfonts}
\usepackage{color}
\usepackage{ulem}
\usepackage{braket}
\usepackage{lineno}

\begin{document}


\title{Inelastic neutron scattering study of crystalline electric field excitations \\
in the caged compounds Nd$\textit{T}_{2}$Zn$_{20}$ (\textit{T} = Co, Rh, and Ir)
}


\author{Rikako Yamamoto}
 \affiliation{
Department of Quantum Matter, Graduate School of Advanced Science and Engineering, Hiroshima University, 
Higashi-Hiroshima, 739-8530, Japan}
\author{Manh Duc Le}
 \affiliation{
ISIS facility, Rutherford Appleton Laboratory, Chilton, Didcot Oxon OX11 0QX, United Kingdom} 
\author{Devashibhai T. Adroja}
 \affiliation{
ISIS facility, Rutherford Appleton Laboratory, Chilton, Didcot Oxon OX11 0QX, United Kingdom}
 \affiliation{
Highly Correlated Matter Research Group, Physics Department, University of Johannesburg, P. O. Box 524, Auckland Park 2006, South Africa}
\author{\\Yasuyuki Shimura}
 \affiliation{
Department of Quantum Matter, Graduate School of Advanced Science and Engineering, Hiroshima University, 
Higashi-Hiroshima, 739-8530, Japan}
\author{Toshiro Takabatake}
\author{Takahiro Onimaru}
 \email{onimaru@hiroshima-u.ac.jp}
 \affiliation{
Department of Quantum Matter, Graduate School of Advanced Science and Engineering, Hiroshima University, 
Higashi-Hiroshima, 739-8530, Japan}



\date{\today}


\begin{abstract}

We have measured crystalline electric field (CEF) excitations of Nd$^{3+}$ ions in the two-channel Kondo lattice candidates Nd$\textit{T}_{2}$Zn$_{20}$ (\textit{T} = Co, Rh, and Ir) by means of inelastic neutron scattering (INS).
In the INS measurements at 5 K, dispersionless excitations were observed at 3.8 and 7.2 meV for \textit{T} = Co, 3.1 and 5.8 meV for \textit{T} = Rh, and 3.0 and 5.3 meV for \textit{T} = Ir.
Analyses of the temperature dependence of the INS spectra confirm that the CEF ground states are the $\Gamma_{6}$ doublet, that is a requisite for manifestation of the magnetic two-channel Kondo effect.
For \textit{T} = Co, a shoulder was observed at 7.7 meV close to the CEF excitation peak centered at 7.2 meV.
The shoulder is attributed to a bound state of the CEF and low-lying optical phonon excitations.
\end{abstract}


\maketitle

\section{Introduction}

Strongly correlated electron systems with caged structures such as filled skutterudite, pyrochlore, and clathrate compounds have attracted much attention \cite{Sato09,Hiroi12,Slack95,Takabatake14}.
In the caged compounds, strong electron-electron and/or electron-phonon interactions are enhanced by entanglement of (quasi-) degenerate degrees of freedom of highly coordinated atoms.
Thereby, they show a variety of phenomena, e.g., heavy fermion behavior, multipole order, unconventional superconductivity, and high-efficient thermoelectricity.
When a rare-earth ion is encapsulated in a highly symmetric cage, 
the crystalline electric field (CEF) effect is weakened and hybridization of the 4$f$ electrons with conduction electrons ($c$--$f$ hybridization) is strengthened in total by the sphere-like arrangement of the caged atoms.

This situation could give rise to anomalous metallic states due to the interaction between low-lying CEF levels.
Among the filled skutterudite compounds, PrOs$_{4}$Sb$_{12}$ shows unconventional superconductivity below $T_{\text{c}}$ = 1.85 K \cite{Bauer02}.
Initially, research interest focused on the issue of how the 4$f^{2}$ electrons would affect the superconducting pair formation. However, the weak CEF splitting hindered a determination of the CEF ground state \cite{Bauer02,Aoki03,Goto04}.
Eventually, it was confirmed that the ground state is a $\Gamma_{1}$ singlet with a first excited $\Gamma_4^{(2)}$ triplet at 0.7 meV above the ground state \cite{Goremychkin04}.
Taking the quasi-degenerate low-lying state into consideration, the 4$f$ collective excitations described as quadrupole excitations could be associated with the heavy-fermion superconducting state \cite{Kuwahara05, Aoki07}.

\begin{figure*}[t]
\begin{center}
\includegraphics[scale=0.9]{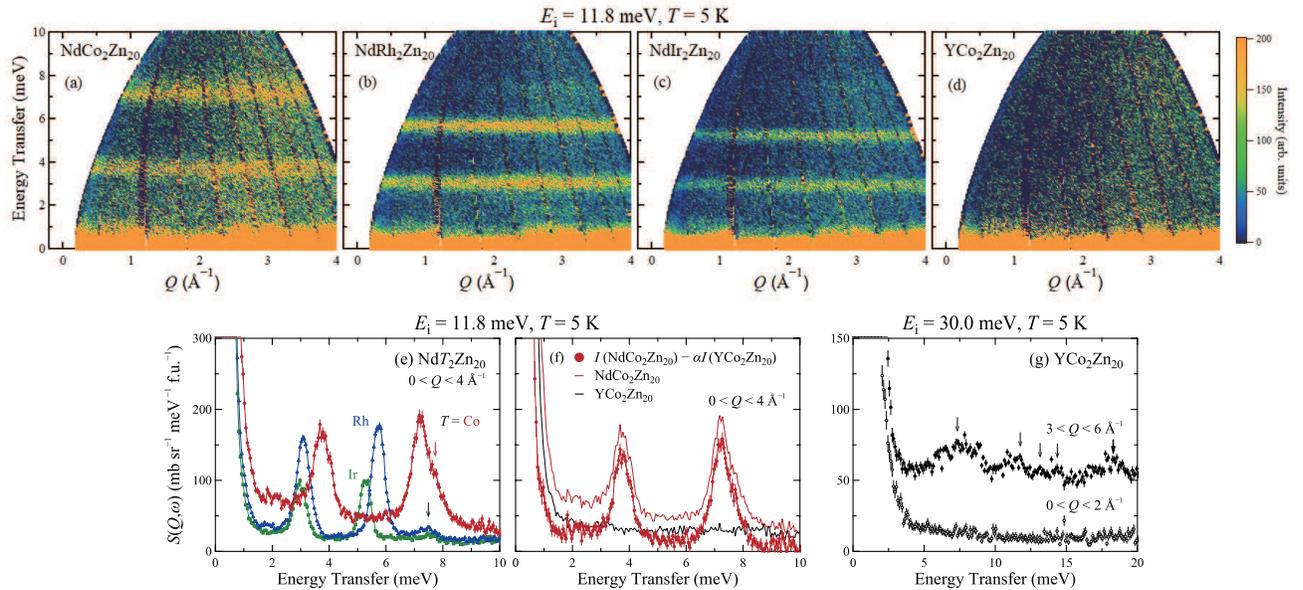}
\caption{(Color online) Color-coded plots of the INS intensity of Nd$\textit{T}_{2}$Zn$_{20}$ for (a) \textit{T} = Co, (b) Rh , and (c) Ir, and (d) nonmagnetic counterpart YCo$_{2}$Zn$_{20}$ as a function of energy and momentum transfers with $E_{\text{i}}$ = 11.8 meV at $T$ = 5 K.
(e) The INS spectra of Nd$\textit{T}_{2}$Zn$_{20}$ for \textit{T} = Co (red), Rh (blue), and Ir (green) as a function of energy transfer obtained by integrating the INS intensity in the ranges of $0 < \bm{Q} < 4\ \AA^{-1}$ with $E_{\text{i}}$ = 11.8 meV at $T$ = 5 K.
(f) The magnetic contribution to the INS intensity of NdCo$_{2}$Zn$_{20}$ (red point) with $E_{\text{i}}$ = 11.8 meV at $T$ = 5 K, estimated by subtracting the INS intensity of YCo$_{2}$Zn$_{20}$ (black line) from that of NdCo$_{2}$Zn$_{20}$ (red line).
A factor $\alpha$ = 1.09 is the ratio of total-scattering crosssection for NdCo$_{2}$Zn$_{20}$ and YCo$_{2}$Zn$_{20}$.
(g) The INS spectra of YCo$_{2}$Zn$_{20}$ at 5 K as a function of energy transfer obtained by integrating the INS intensity in the ranges of $0 < \bm{Q} < 2\ \AA^{-1}$ and $3 < \bm{Q} < 6\ \AA^{-1}$ measured with $E_{\text{i}}$ = 30.0 meV.
}
\label{fig01}
\end{center}
\end{figure*}

For another family of compounds
$R\textit{T}_{2}X_{20}$ ($R$: rare-earth element, \textit{T}: transition metal, $X$ = Al, Zn, Cd, and Mg), transport and magnetic properties have been extensively studied in the past decade. 
They crystallize in the cubic CeCr$_{2}$Al$_{20}$-type structure with the space group of $Fd\bar{3}m$ (No. 227, $O_{7}^{h}$) and an $R^{3+}$ ion with the cubic $T_{d}$ point group is encapsulated in a Frank-Kasper cage formed with sixteen $X$ atoms \cite{Nasch97}.
In non-Kramers $\Gamma_{3}$ doublet systems with 4$f^2$ configuration, e.g., Pr$\textit{T}_{2}$Zn$_{20}$ (\textit{T} = Rh and Ir) and Pr$\textit{T}_{2}$Al$_{20}$ (\textit{T} = Ti and V), superconducting transitions manifest themselves in the presence of quadrupole order \cite{Onimaru10,Onimaru11,Onimaru12b,Sakai11,Sakai12,Tsujimoto14}.
Non-Fermi liquid (NFL) behaviors of the electrical resistivity $\rho(T)$, the magnetic specific heat $C_{\text{m}}$, and the elastic constant are consistent with the scenario of the two-channel (quadrupole) Kondo effect, where the local quadrupole is over-screened by the two equivalent conduction bands \cite{Onimaru16b,Tsuruta15,Yamane18a,Yanagisawa19,Worl22}.

In isostructural Nd$\textit{T}_{2}X_{20}$ with 4$\textit{f}^{\,3}$ configuration, on the other hand, ferromagnetic (FM) or antiferromagnetic (AFM) transitions were observed depending on the combination of \textit{T} and $X$ \cite{Isikawa13,Wakiya15,Namiki16b,Yamane17,Yamamoto19}.
They are proposed as candidates exhibiting magnetic two-channel Kondo effect, i.e., the magnetic analogue of the quadrupole Kondo effect, by a numerical renormalization group calculation with a seven-orbital impurity Anderson model \cite{Hotta17}.
The theoretical calculation proves the residual entropy of 0.5$R$ln2, that is a characteristic of the two-channel Kondo effect, in a wide parameter range when the CEF ground state of the Nd ion is the $\Gamma_{6}$ doublet.
Among the Nd$\textit{T}_{2}X_{20}$ family, NdCo$_{2}$Zn$_{20}$ with the smallest lattice parameter is a promising candidate to exhibit the two-channel Kondo effect since the strong $c$--$f$ hybridization is expected by positive chemical pressure effect \cite{Yamamoto19}.
NdCo$_{2}$Zn$_{20}$ shows the AFM transition at $T_{\text{N}}$ = 0.53 K, above which $\rho(T)$ exhibits a downward convex curvature up to 4 K.
This anomalous $\rho(T)$ can be fitted using a theory on the basis of the two-channel Anderson lattice model \cite{Tsuruta15}.
In addition, the magnetic entropy $S_{\rm m}$ at $T_{\text{N}}$ is only 50\% of $R$ln2 expected for the $\Gamma_{6}$ doublet.
These results corroborate the formation of the two-channel Kondo lattice.

However, the CEF level scheme of NdCo$_{2}$Zn$_{20}$ is still controversial as described below.
A Schottky anomaly of $C_{\text{m}}$ at around 13 K is fitted with a CEF level scheme of $\Gamma_{6}(0)-\Gamma_{8}^{(1)}(43\,\text{K})-\Gamma_{8}^{(2)}(72\,\text{K})$, 
and $\chi(T)$ and isothermal magnetization $M(B)$ data are reproduced as well.
On the other hand, the elastic constant $C_{44}$ displays pronounced softening on cooling from 40 to 4.2 K \cite{Umeno20}.
This softening can not be explained by the isolated $\Gamma_{6}$ doublet ground state with no quadrupolar degrees of freedom.
The softening would be explained if the CEF ground state of the $\Gamma_{8}$ quartet or the $\Gamma_{6}$ doublet with
a low-lying excited state, e.g., $\Gamma_{6}(0)-\Gamma_{8}$(11\,\text{K}).

The aims of this study are to reveal whether the CEF ground states of NdCo$_{2}$Zn$_{20}$ and the isovalent compounds Nd$\textit{T}_{2}$Zn$_{20}$ (\textit{T} $=$ Rh: $T_{\text{N}}$ = 0.94 K, and Ir: 0.65 K \cite{Yamane17,Yamamoto19}) are the isolated $\Gamma_{6}$ doublet and how the CEF excitations are associated with the low-lying phonon contribution arising from the caged structure.
For this purpose, we have performed powder inelastic neutron scattering (INS) measurements
In addition,  a nonmagnetic counterpart YCo$_{2}$Zn$_{20}$ was measured to extract the phonon excitations.

\section{Experimental procedure}

Single crystals of NdCo$_{2}$Zn$_{20}$ were synthesized by the Zn self-flux method, as described in previous reports \cite{Jia09,Yamamoto19}.
Polycrystalline samples of Nd$\textit{T}_{2}$Zn$_{20}$ (\textit{T} = Rh and Ir) and YCo$_{2}$Zn$_{20}$ were prepared by the melt-growth method \cite{Yamane17,Yamamoto19}.
For the INS measurements, we used powdered samples of 3.4, 7.6, 8.5, and 4.8 g for \textit{T} = Co, Rh, Ir, and YCo$_{2}$Zn$_{20}$, respectively.
The INS measurements were carried out with the time-of-flight (TOF) spectrometer MARI installed at the ISIS Facility, Rutherford Appleton Laboratory.
The incident neutron energies were selected as $E_{\text{i}}$ = 6.2, 9.9, 11.8, 30.0, and 100 meV.
The powered samples packed in an aluminum can were cooled down to 5 K with a top loading closed cycle refrigerator.
INS spectra with energy and momentum transfers were obtained from TOF signals using the Mantid software \cite{Arnold14}.
The spectral intensity was calibrated by using the crosssection of a standard vanadium sample.

\section{Results and Discussion}

Figures \ref{fig01}(a)--(d) are color-coded plots of the INS intensity of Nd$\textit{T}_{2}$Zn$_{20}$ for \textit{T} = Co, Rh, and Ir and YCo$_{2}$Zn$_{20}$ at $T$ = 5 K as functions of energy and momentum transfers with $E_{\text{i}}$ = 11.8 meV.
There are two dispersionless excitations at 3.8 and 7.2 meV for \textit{T} = Co, 3.1 and 5.8 meV for \textit{T} = Rh, and 3.0 and 5.3 meV for \textit{T} = Ir, respectively, whereas such excitations are absent in YCo$_{2}$Zn$_{20}$.
Therefore, the dispersionless excitations of Nd$\textit{T}_{2}$Zn$_{20}$ are ascribed to transitions between the CEF levels of the Nd$^{3+}$ ions.
No further magnetic excitation was observed in the higher energy range of $E$ $\ge$ 10 meV by the measurements with $E_{\text{i}}$ = 30.0 and 100 meV.
Thereby, the overall energy scale of the CEF excitations in the Nd ions is within 10 meV.

Figure \ref{fig01}(e) displays the INS spectra of Nd$\textit{T}_{2}$Zn$_{20}$ for \textit{T} = Co (red circles), Rh (blue triangles), and Ir (green boxes) at 5 K obtained by integrating the INS data in the momentum transfer ranges of $0 < \bm{Q} < 4\ \AA^{-1}$.
The data include not only the magnetic excitations but also phonon contributions, since high intensity excitations are seen in the spectrum of the nonmagnetic YCo$_{2}$Zn$_{20}$ for $\bm{Q} > 2$ ${\AA}^{-1}$.
Note that the intensity of \textit{T} = Ir is smaller than that of \textit{T} = Co and Rh.
It is ascribed to the larger neutron absorption crosssection of the natural Ir element than that for Co and Rh.
The CEF excitation energies of 3.8 and 7.2 meV for \textit{T} = Co are higher than 3.1 and 5.8 meV for \textit{T} = Rh and 3.0 and 5.3 meV for \textit{T} = Ir, respectively.
This decreasing sequence in energy going from \textit{T} = Co to Ir is consistent with the weakening of CEF upon increasing the lattice parameter $a$ from 14.110 $\AA$ to 14.425 $\AA$ and 14.426 $\AA$ \cite{Yamane17,Yamamoto19}.

The background intensity in the spectra of \textit{T} = Co is apparently higher than that of \textit{T} = Rh and Ir.
Thus, the magnetic contribution to the total intensity of NdCo$_{2}$Zn$_{20}$ was extracted by subtracting the INS intensity of YCo$_{2}$Zn$_{20}$ as the phonon contribution.
In Fig. \ref{fig01}(f), the total intensity of NdCo$_{2}$Zn$_{20}$ (red line) and YCo$_{2}$Zn$_{20}$ (black line) are displayed.
Here, the intensity of YCo$_{2}$Zn$_{20}$ was multiplied by a factor of $\alpha$ = 1.09 to fit the total-scattering crosssection to that of NdCo$_{2}$Zn$_{20}$.
Eventually, the magnetic contribution to the intensity of NdCo$_{2}$Zn$_{20}$ is shown with the (red) circles.
The background level of NdCo$_{2}$Zn$_{20}$ is comparable to those for \textit{T} = Rh and Ir shown in Fig. \ref{fig01}(e).
In fact, the phonon contribution can be recognized from the color-coded plot of the INS intensity for YCo$_{2}$Zn$_{20}$ in Fig. \ref{fig01}(d).
Fig. \ref{fig01}(g) shows the INS intensity obtained by integrating the INS data for $3 < \bm{Q} < 6\ \AA^{-1}$ (closed circles) and $0 < \bm{Q} < 2\ \AA^{-1}$ (open circles), respectively.
As indicated with the arrows, the phonon density of states is recognized in the spectrum for $3 < \bm{Q} < 6\ \AA^{-1}$, whereas the intensity is suppressed in the spectrum for $0 < \bm{Q} < 2\ \AA^{-1}$.

The phonon contributions were detected as well at around 7.5 meV in the INS spectra of Nd$\textit{T}_{2}$Zn$_{20}$ for \textit{T} = Rh and Ir as shown with the (black) arrow in Fig. \ref{fig01}(e).
The momentum transfer dependence of the INS intensity is likely consistent with the phonon dispersion relations of the isostructural compounds Pr$\textit{T}_{2}$Zn$_{20}$ (\textit{T} = Rh and Ir), yielding that the optical phonon modes at 6$-$7 meV due to the vibrations of Zn atoms \cite{Wakiya21}.
On the other hand, for \textit{T} = Co, near the CEF excitation peak centered at 7.2 meV, a shoulder appears at 7.7 meV as marked with the (red) arrow in Fig. \ref{fig01}(e).
A possible origin of the shoulder at 7.7 meV is discussed later.
Concerning the peak width of the excitations, the excitation peaks for \textit{T} = Co are broader than those for \textit{T} = Rh and Ir.
One explanation of the broadening of the peaks is the enhanced $c$--$f$ hybridization effect \cite{Fulde85}.
The $c$--$f$ hybridization effect for \textit{T} = Co is probably the strongest among them since the lattice parameter is the smallest. 
Another possibility is that the CEF excitations are still dispersive due to short-range magnetic correlations above $T_{\text{N}}$.
Otherwise, the broadening is ascribed to a possible coupling between the CEF and optical phonon excitations as discussed later.
It is thus necessary to measure the CEF excitations in single crystals.

\begin{figure}
\begin{center}
\includegraphics[scale=0.4]{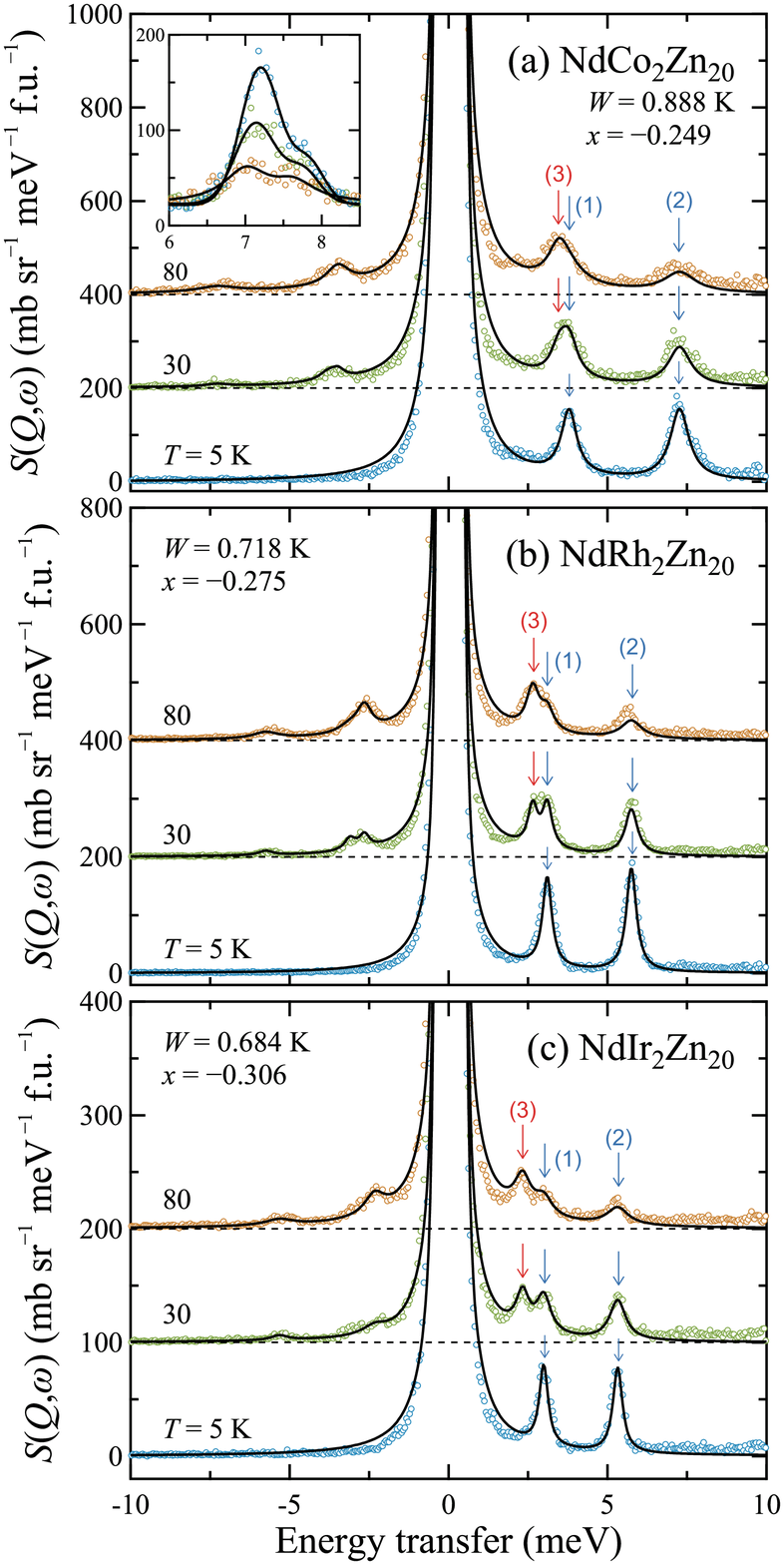}
\caption{(Color online) INS spectra as a function of energy transfer of Nd$\textit{T}_{2}$Zn$_{20}$ for (a) $\textit{T}$ = Co, (b) Rh, and (c) Ir at $T$ = 5, 30, and 80 K.
The data are obtained by integrating the INS intensity in the ranges of $0 < \bm{Q} < 2\ \AA^{-1}$.
The solid lines are the fits based on the cubic CEF model as described in the text.
The data at $T$ = 30 and 80 K are vertically shifted for clarify.
}
\label{fig02}
\end{center}
\end{figure}

Figures \ref{fig02}(a)--(c) exhibit the INS spectra of Nd$\textit{T}_{2}$Zn$_{20}$ for \textit{T} = Co, Rh, and Ir at $T$ = 5, 30, and 80 K.
The data are obtained by integrating the INS intensity in the ranges of $0 < \bm{Q} < 2\ \AA^{-1}$ to ignore the phonon contributions.
First, we demonstrate the temperature dependent INS spectra for \textit{T} = Ir since the the excitation peaks are well separated at each temperature as shown in Fig. \ref{fig02}(c).
Two peaks at 3.0 and 5.3 meV observed at $T$ = 5 K as labeled (1) and (2) are attributed to the CEF excitations from the ground state to the excited states.
With increasing the temperature to 30 K, the intensity of both the peaks decreases.
This decrease  in the intensity is caused by the depopulation of the ground state on warming.
Instead, an additional peak appears at 2.3 meV at $T$ = 30 K as labeled (3), and it develops at 80 K, yielding that the first excited state is thermally populated.

Similar temperature dependences of the INS spectra were observed for \textit{T} = Co and Rh as shown in Figs. \ref{fig02}(a) and \ref{fig02}(b): The two peaks of at 3.8 (3.1) and at 7.2 (5.8) meV for \textit{T} = Co (Rh) due to the transitions from the ground state to the first and the second excited states appear at 5 K and the additional peak at 3.3 (2.6) meV arising from the transition between the first and second excited states develops on warming for $T$ $\geq$ 30 K.

\begin{figure}[t]
\begin{center}
\includegraphics[scale=0.92]{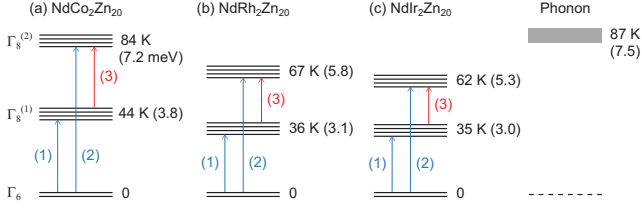}
\caption{(Color online) CEF level schemes of Nd$^{3+}$ ions in Nd$\textit{T}_{2}$Zn$_{20}$ for \textit{T} = Co, Rh, and Ir, and the optical phonon density of states at around 87 K.
The values in parentheses are in unit of meV.
}
\label{fig03}
\end{center}
\end{figure}

The temperature dependences of the INS spectra were analyzed on the basis of the cubic CEF model.
With the cubic point group of the Nd site, the CEF split the tenfold-degenerate multiplet of the $J = 9/2$ state into a $\Gamma_{6}$ doublet and two $\Gamma_{8}$ quartets. 
Here, the transitions between the three multiplets are permitted, whereas the transition matrix elements strongly depend on the CEF parameters \cite{Birgeneau72,Fulde85}. 
The CEF Hamiltonian for the cubic point group of $T_{d}$ is written in terms of CEF parameters $x$ and $W$, and the Stevens equality operators $O_{n}^{m}$ as follows \cite{Lea62}, 
\begin{equation}
\mathcal{H}_{\textit{CEF}} = W\left[\frac{x}{60}(O_{4}^{0}+5O_{4}^{4})+\frac{1-|x|}{2520}(O_{6}^{0}-21O_{6}^{4})\right].
\end{equation}
First, the initial parameters were set to take into account the ratio of excitation energies at 5 K, and the CEF ground state was selected to be either the $\Gamma_{6}$ doublet or the $\Gamma_{8}$ quartet.

Within the dipole approximation, the magnetic crosssection is represented as \cite{Squires96}
\begin{equation*}
\frac{d^{2}\sigma}{d\Omega dE_{f}} = (\gamma r_{0})^{2}\frac{k_{f}}{k_{i}}\Bigl\{\frac{1}{2}g_{J} f(Q)\Bigr\}^{2} \sum_{\alpha,\,\beta}(\delta_{\alpha,\,\beta}-\hat{Q}_{\alpha}\hat{Q}_{\beta})S^{\alpha\,\beta}(Q,\varepsilon),
\end{equation*}
where
\begin{equation}
S^{\alpha\,\beta}(Q,\varepsilon) = \sum_{\lambda,\,\lambda'}p_{\lambda}|\bra{\lambda}J_{\alpha}(Q)\ket{\lambda'}\bra{\lambda'}J_{\beta}(Q)\ket{\lambda}|\delta(\varepsilon-E_{\lambda'}+E_{\lambda}).
\end{equation}
Here, $Q$ and $\varepsilon$ are momentum and energy transfer.
$\gamma$ = 1.913, $r_{0} = 2.818 \times 10^{-15}$ m and  $g_{J}$ are the neutron gyromagnetic ratio,  the classical electron radius, and Land$\acute{\text{e}}$ $g$-factor.
The magnetic form factor of a single Nd$^{3+}$ ion is given by $f(Q)$.
$k_{i}$ and $k_{f}$ are initial and final neutron wave vectors.
An eigenfunction $\ket{\lambda}$ has the energy of $E_{\lambda}$, and $J_{\alpha}$ and $J_{\beta}$ ($\alpha$, $\beta$ $=$ $x$, $y$, $z$) are the components of the total angular moment.
$p_{\lambda}$ is the probability that the neutron is initially in the state $\lambda$.

In the analysis, a least squared fitting was performed by applying the Lorentzian lineshape to the inelastic peaks and a pseudo-Voigt lineshape to the elastic peak at $E$ = 0, respectively. 

The fits are shown with the solid lines in Fig. \ref{fig02} and the proposed CEF levels of the Nd$^{3+}$ ions for \textit{T} = Co, Rh, and Ir are illustrated in Fig. \ref{fig03}.
The CEF level schemes are as follows: $\Gamma_{6}(0\,\text{K})-\Gamma_{8}^{(1)}(44\,\text{K})-\Gamma_{8}^{(2)}(84\,\text{K})$ for \textit{T} = Co,  $\Gamma_{6}(0\,\text{K})-\Gamma_{8}^{(1)}(36\,\text{K})-\Gamma_{8}^{(2)}(67\,\text{K})$ for \textit{T} = Rh, and  $\Gamma_{6}(0\,\text{K})-\Gamma_{8}^{(1)}(35\,\text{K})-\Gamma_{8}^{(2)}(62\,\text{K})$ for \textit{T} = Ir.
These level schemes are in good agreement with those proposed from the analyses of the specific heat data \cite{Yamane17,Yamamoto19}.
Therefore, we conclude that the CEF ground states of the Nd$^{3+}$ ions in the three compounds are the $\Gamma_{6}$ doublets isolated from the first excited $\Gamma_{8}^{(1)}$ quartets by 35--44 K. 
Therefore, the three compounds are the ideal platform to verify the two-channel Kondo effect.

Next, we discuss the possible coupling between the CEF and low-lying phonon excitations in NdCo$_{2}$Zn$_{20}$.
In general, because of the weak coupling between the CEF and phonon excitations, the dispersionless CEF and dispersive phonon excitations could be independently measured. 
However, it was suggested that a CEF-phonon bound state is realized in some rare-earth based compounds such as CeAl$_{2}$ \cite{Thalmeier82,Thalmeier84}, YbPO$_{4}$ \cite{Loong99}, $R$Cu$_{2}$ ($R$ = Ce and Nd) \cite{Loewenhaupt88,Loewenhaupt90,Hense04}, Ce$\textit{T}$Al$_{3}$ (\textit{T} = Cu and Au) \cite{Adroja12,Cermak19}, and CeCuGa$_{3}$ \cite{Anand21}.
In the case of CeAl$_{2}$, a strong magnetoelastic coupling of the transverse strain components to the CEF states of Ce$^{3+}$ gives rise to a pronounced softening of the $C_{44}$ elastic constant \cite{Luthi79}.

Coming back to the INS spectrum in NdCo$_{2}$Zn$_{20}$, 
the shoulder at 7.7 meV may probably be attributed to the bound state of the CEF excitation and the optical phonon modes for the following reasons.
(1) The shoulder was observed in the INS spectra for $0 < \bm{Q} < 2\ \AA^{-1}$.
(2) The intensity at the shoulder is temperature dependent as shown in the inset of Fig. \ref{fig02}(a).
(3) The intensity is much higher than that of the phonon excitations at 7.5 meV for \textit{T} = Rh and Ir.
(4) The CEF excitation energy of 7.2 meV (84 K) is close to 7.5 meV (87 K) of the optical phonon branches.
Thereby, the coupling of the CEF and the optical phonon modes may give rise to the broadening of the CEF excitation peaks.
Here, it is noted that the $Q$-dependence of the INS intensity in NdCo$_{2}$Zn$_{20}$ cannot be explained by considering the form factor of a Nd$^{3+}$ ion and the INS intensity of YCo$_{2}$Zn$_{20}$ as the phonon contribution.
This discrepancy may be ascribed to not only difference of the lattice parameters but also the CEF-phonon coupling.
In addition, as described in the introduction, the $C_{44}$ elastic constant exhibits the pronounced softening on cooling from 40 to 4.2 K.
Since the contribution of the quadrupolar degrees of freedom in the low-lying CEF levels is ruled out, the magnetoelastic coupling may give rise to the elastic softening \cite{Umeno20}. 
To unveil the magnetoelastic coupling, it is necessary to measure the phonon dispersion relations by means of 
Raman spectroscopy, inelastic x-ray or neutron scattering.

\section{Summary}

The CEF level schemes of the Nd$^{3+}$ ions in the caged compounds Nd$\textit{T}_{2}$Zn$_{20}$ (\textit{T} = Co, Rh, and Ir) have been investigated by means of the TOF inelastic neutron scattering experiments.
By analyzing the temperature dependences of the INS spectra from 5 to 80 K, we determined the CEF level schemes to be composed of the $\Gamma_{6}$ doublet ground states and two $\Gamma_{8}$ excited states.
Thereby, the three compounds are good candidates to verify the two-channel Kondo effect.
For \textit{T} = Co, a shoulder was observed at 7.7 meV near the CEF excitation peak centered at 7.2 meV.
We ascribe this shoulder to a bound state due to the strong CEF-phonon coupling.


\begin{acknowledgments}

Experiments at the ISIS Neutron and Muon Source were supported by a beamtime allocation RB1920299 from the Science and Technology Facilities Council. Data is available here: https://doi.org/10.5286/ISIS.E.RB19202979.
We would like to thank Y. Kusanose and T. Guidi for assistance of the inelastic neutron scattering measurements with the TOF neutron spectrometer MARI at ISIS.
We  also thank fruitful discussion with H. Harima, K. Kusunose, K. Umeo and S. Tsutsui.
TO and DTA acknowledge the Royal Society of London for the  International Exchanges funding ref: IEC\textbackslash R3\textbackslash 203022.
This work was financially supported by MEXT/JSPS KAKENHI Grant Numbers JP26707017, JP15H05886 and JP15K21732 (J-Physics), and JP18H01182, Japan. 

\end{acknowledgments}

\bibliography{20230120_NdTr2Zn20_ins_rikako_12}

\end{document}